# Bell's Theorem and Nonlinear Systems

Louis Vervoort

*LPMC, Ecole Normale Supérieure, 24 Rue Lhomond, 75231 Paris cedex 5,*

*and IEF(\*), Université Paris-Sud, Bât. 220, 91405 Orsay cedex, France.*

Abstract

For all Einstein-Podolsky-Rosen-type experiments on deterministic systems the Bell inequality holds, unless non-local interactions exist between certain parts of the setup. Here we show that in nonlinear systems the Bell inequality can be violated by non-local effects that are *arbitrarily* weak. Then we show that the quantum result of the existing Einstein-Podolsky-Rosen-type experiments can be reproduced within deterministic models that include arbitrarily weak non-local effects.

(*) Present address. E-mail: Louis.Vervoort@ief.u-psud.fr

Since the early days of quantum mechanics people have wondered whether this remarkably efficient theory could be made deterministic, like classical theories. In 1965 J.S. Bell [1] proved that there is a fundamental difference between quantum mechanics and such classical theories: he showed, in the case of the Einstein-Podolsky-Rosen-Bohm (EPRB) experiment, that all "local hidden-variable theories (HVTs)" predict different results as quantum mechanics. (The term "local" is used here in the original sense of ref.[1], to indicate interactions that have only a limited range in space[1].) Shortly after Bell's work, several authors derived a more general version of Bell's theorem, stating that basically *any* local realistic theory (deterministic or not) is in conflict with quantum mechanics [2]. If quantum theory is correct this would imply that quantum systems cannot, in general, be considered as local. The last decades several groups have performed experiments that come closer and closer to the ideal EPRB scheme, and as is well known the quantum result proved correct. Thus a first series of experiments (both with static [3,4] and periodically switching [5] analyser settings) convinced physicists that "local realism" is refuted, and in particular that quantum systems cannot be considered as both local and deterministic. A very recent experiment with randomly changed settings [6] would go one step further by proving that HVTs reproducing quantum mechanics are not only non-local, but need to invoke superluminal interactions. It should be pointed out that some loopholes were left open by all existing experiments [3,7,8]. Nevertheless one generally accepted viewpoint in the physics community seems to be that if a deterministic HVT wants to describe nature, it must include non-local effects.

Here we consider hidden variable models for the EPRB experiment that include such non-local interactions (in the sense of Bell[1]) between distant parts of the experimental setup.

---

[1] According to ref. [1] locality means: the setting of a measuring device does not influence the result obtained with another remote measuring device, nor does it influence the way in which particles are emitted by a distant source. As in ref. [1], we do *not* use the term "local" here in the broader sense to designate interactions with a (sub)luminal speed of propagation.

If these interactions are due to the known (electromagnetic and gravitational) long-range fields, which have a rapidly decreasing intensity at large distance, one may expect them to have a negligible influence. This is the starting point of Bell's classic analysis. But this seems not *necessarily* true if a complex reality would underlie the quantum level, and if some hidden variables would exhibit nonlinear, eventually chaotic, behaviour. It is well known that in chaotic systems even weak fluctuations (for instance induced by long-range interactions) can be dynamically enhanced, due to an exponential sensitivity upon initial conditions. Moreover, one could notice that in some chaotic systems phenomena with a certain "non-local" character have been observed [9]. Guided by these ideas we consider here the possibility that some hidden variables $\lambda$ are a function of both the left and right analyser positions (denoted by a, respectively b) under the condition that (i) the $\lambda$ evolve in a nonlinear manner *and* (ii) that the dependence of the $\lambda$ on (a, b) is weak. This means that one can write $\lambda = \lambda_L + \lambda_{NL}(a,b)$, with $\lambda_L$ a purely "local" contribution and $\lambda_{NL}$ a small non-local correction that depends on a and b. Hence we use the term "weak non-local effect". The main result of this article is that the quantum result of the EPRB experiments [3-5] can be reproduced within HVTs that include only *arbitrarily* weak non-local (but Lorentz-invariant) effects. Such effects may have escaped from experimental investigation. This result holds for all but one of the existing tests, which use static or periodic polarizer settings. Therefore it emphasises the importance of tests with rapid and random switching, like in a very recent experiment [6].

We first succinctly recall Bell's analysis as applied to the optical version of the EPRB experiment [1,3]. Pairs of photons are emitted from a source, one photon (say I) going to the left, the other (II) to the right. Photon I (II) interacts with polarizer $P_I$ ($P_{II}$) which has its polarization axis in direction a (b). In this idealised experiment one measures for each emitted pair the dichotomic properties $A_I(a)$ and $A_{II}(b)$, defined as follows: $A_I(a) = +(-)$ 1 if photon I passes (not) through polarizer $P_I$ which has its polarization axis in the direction a; and analogous for $A_{II}(b)$. If the photon pairs are emitted in a well-defined polarization state (analogous to the spin singlet state for electrons), then quantum mechanics predicts for the average value of the product $A_I.A_{II}$: $\langle A_I(a).A_{II}(b)\rangle = \cos 2(a-b)$. Within a local HVT this mean value would be given by:

$$M(a,b) = \int d\bar{\lambda} \cdot \rho(\bar{\lambda}) \cdot A_I(a,\bar{\lambda}) \cdot A_{II}(b,\bar{\lambda}) \quad , \qquad (1)$$

with $\rho(\bar{\lambda})$ the probability distribution of the hidden-variable set $\bar{\lambda}$. Such HVTs are "local" (see footnote (1)) in the sense that $\rho$ does not depend on a nor b (in the following we will focus our attention on this element of the locality condition) and that $A_I$ is only a function of a and not of b (vice versa for $A_{II}$). The Bell inequality states then that the quantity $S = M(a,b) - M(a,b') + M(a',b) + M(a',b')$ is bounded by $\pm 2$ for any set of "analyser knob" positions (a, b, a', b'). But the experimental results [3-6] violate this inequality for some sets of positions, and are in agreement with the quantum prediction.

Bell's locality condition is a very reasonable assumption, but in experiments with static setups [3,4] it does not *necessarily* hold. In such experiments an exchange of Lorentz-invariant signals between the subsystems is possible, implying that $A_{I(II)}$ can become, in principle, also a function of b (a), or $\rho(\bar{\lambda})$ a function of a and b. Of course, if the subsystems are far apart it seems only natural to neglect this non-local dependence, even in static experiments. But as argued in the introduction, if a complex reality would underlie the quantum level and if the hidden photon dynamics would be nonlinear, then it seems worthwhile to study even weak fluctuations of the form $\bar{\lambda} = \bar{\lambda}(a,b)$ or more precisely $\rho = \rho(\bar{\lambda},a,b)$. These ideas can be put in a slightly different context: in order to measure S in static experiments one has to perform four successive experiments $M_i$ (i = 1,...,4) with different analyser settings, and:

$$S = M_1(a,b) - M_2(a,b') + M_3(a',b) + M_4(a',b'). \qquad (2)$$

In a deterministic picture the values of the initial conditions ($\bar{\lambda}$) which determine $A_{I,II}$ (and thus M) will not be *exactly* the same in the four experiments. This may be due to small stochastic fluctuations in the preparation, or to a (weak) long-range interaction between the analysers and the particle source. Note that both phenomena can formally lead to $\rho = \rho(\bar{\lambda},a,b)$. If the $\bar{\lambda}$ pertain to a non-linear system, and if one considers distribution functions of the type $\rho(\bar{\lambda}) = (1/N) \Sigma_j \delta(\bar{\lambda} - \bar{\lambda}_j)$, $j = 1,..., N$, then even small differences in the $\bar{\lambda}_j$ and thus in $\rho(\bar{\lambda})$ may lead to completely different values of M, as will be shown hereafter.

To quantify these ideas we have studied a classical model system which is known to possess a rich dynamical behaviour, namely a compass in a fixed plus a rotating magnetic field [10]. (The dynamics of this system, eq. (3), may give us inspiration about how the hidden photon dynamics might formally behave.) Both the rotating field $B_r$ and the fixed field $B_f$ lie in the horizontal plane of the compass needle, and $B_r$ turns with frequency $\omega$ around the centre of suspension of the compass like the pointer of a clock. Depending on the values of $B_r$, $B_f$ and $\omega$ the compass will oscillate around its centre in a chaotic or a regular (periodic) manner. If the compass has an inertial moment J and a magnetic moment $\mu$, one readily finds that the angle $\theta$ between the compass and $B_f$ is governed by the nonlinear equation:

$$\ddot{\theta} + \alpha \dot{\theta} = - x \sin\theta - P \sin(\theta - t). \tag{3}$$

Here $\alpha \dot{\theta}$ is a frictional term and $x = \mu B_f / J\omega^2$ and $P = \mu B_r / J\omega^2$ are introduced to obtain an equation with only dimensionless parameters and with $1/\omega$ as time unit. In the numerical calculations we take $\alpha = 0.174$, $P = 0.335$ and $x \in [0.1600 ; 0.2321]$, as for these values (3) is known to reproduce the experimental results. By increasing the bifurcation parameter x (in the experiments $B_f$) this system goes through a subharmonic cascade and becomes chaotic for x larger than about 0.2285. For such x-values the system depends sensitively on initial conditions: while in the regular regime two trajectories with slightly different initial conditions are indistinguishable, they diverge in the chaotic regime.

Now we construct an analogue of the EPRB experiment [1] based on this system. Instead of photons the experiment involves measurements on two identical, separated, compasses I and II, each in a fixed plus a rotating field. (We could have constructed an explicit HVT for photons via eq. (3), by considering θ as a hidden variable characterising the photon and determining the measurement result of $A_{I,II}$ in the way defined below. But this is not very useful.) We will suppose that $B_r$ is the same for I and II, but that $B_f$, i.e. x, can be different. One measures simultaneously the orientations $θ_I$ and $θ_{II}$ of the compasses at a given measuring time $t_m$. The corresponding dichotomic properties $A_{I,II}$ are defined as follows: $A_{I,II} = + 1$ if $| θ_{I,II}(t_m) | < Δ$, and else -1. $A_I$ and $A_{II}$ clearly depend on the "analyser knob" variable x. The role of the hidden variables $\bar{λ}$ is attributed to the initial conditions ($θ(0); \dot{θ}(0)$), which we assume to be the same for both compasses (due to a "correlation" at t = 0). If one considers an ensemble of such correlated systems with probability density $ρ(\bar{λ})$ for the initial values, then one can calculate the mean value of the product $A_I(a).A_{II}(b)$ over the ensemble. This mean value is given by (1) if x = a for I and x = b for II.

Clearly the Bell inequality holds for this dichotomic experiment on two deterministic, local, subsystems. Let us calculate S as given by (2) for (a, b, a', b') = (0.160, 0.170, 0.230, 0.232). In Table I the compass orientations at $t_m$ = 100 are given for these parameter values, calculated for $\bar{λ}$ = (0.6; 0). If one supposes that this is the only $\bar{λ}$ in the ensemble (i. e. $ρ(\bar{λ}) = δ(\bar{λ})$) and that Δ = 0.3, then one immediately finds that S = 2 for this experiment, in accordance with the Bell inequality (if $ρ(\bar{λ}) = δ(\bar{λ})$ it is trivial to prove that S must be equal to +2 or -2). Table I also gives the compass orientations for $\bar{λ}$ = (0.6; $10^{-3}$), calculated for the four x-values. As expected, in the regular regime (x = a or b) these initial values lead to the same θ-values as in the case of (0.6; 0), but in the chaotic regime (x = a' or b') one finds very different θ-values. For a' property A changes its sign.

As stated above, in static EPRB-type experiments measurement of S involves four successive experiments. We now make the assumption that the values of the initial conditions (more precisely ρ($\bar{\lambda}$)) are not exactly the same during the four measurements (for instance as a consequence of an interaction between analysers and "particle" source). To be definite, consider again above model system and suppose that during measurement of $M_1$(a,b), $M_2$(a,b') and $M_3$(a',b) $\bar{\lambda}$ = (0.6; 0), and that for $M_4$(a',b') $\bar{\lambda}$ = (0.6; $10^{-3}$). If one calculates now S as above one finds with the values of Table I that it is equal to 4 in this experiment. Clearly, this violation of the Bell inequality in a deterministic system is no surprise if one realises that the assumption of different $\bar{\lambda}$ during different measurements amounts to the introduction of a non-local effect (even if the subsystems would not interact). Nevertheless it is intriguing that such a drastic effect can be induced by differences in initial conditions between the four experiments that are very small. There is no difficulty in finding combinations of apparatus settings for which differences in $\bar{\lambda}$ that are orders of magnitude smaller than $10^{-3}$ lead to a violation of the Bell-inequality. For instance for (0.16007, 0.16008, 0.16009, 0.230069) and Δ = 0.001, initial values during measurement of $M_2$ that are equal to (0.6 ; $10^{-5}$) in stead of (0.6 ; 0) lead again to S = 4. As a matter of fact, by sufficiently decreasing $\Delta$ or by sufficiently increasing the measuring time $t_m$ one can violate the Bell inequality for an *arbitrarily small* fluctuation in the initial conditions [11].

|  | $\dot{\theta}(0) = 0$ | $\dot{\theta}(0) = 1.\ 10^{-3}$ |
|---|---|---|
| a = 0.160 | 0.01 | 0.01 |
| b = 0.170 | -0.29 | -0.29 |
| a' = 0.230 | -0.19 | -0.45 |
| b' = 0.232 | 1.10 | 0.86 |

TABLE I. Angle θ (in radians) as calculated with equation (3), at $t_m$ = 100, for 4 x-values and for 2 slightly different initial values ($\theta(0)$; $\dot{\theta}(0)$) = (0.6; 0) and (0.6; $10^{-3}$).

The conclusion we draw from this example is that it proves that the Bell inequality can be violated in deterministic systems, for given static analyser settings, by non-local (but Lorentz-invariant) effects that are vanishingly small. Such vanishingly weak non-local corrections are of course much more "probable" than strong effects. Clearly, if one looks for a HVT that can reproduce the quantum result of the EPRB-experiments [3-6] it should be able to predict the cos2(a-b) dependence of the correlation function. We will now demonstrate that such HVTs can be constructed; it seems a priori clear that these have to include non-local effects. The main thesis of the present article is that the quantum result of the static and periodic experiments [3-5] can be reproduced by vanishingly weak non-local (but Lorentz-invariant) effects.

Consider again the above EPRB-type experiment (on pairs of compasses, or photons, of which the dynamics is governed by eq. (3)), and suppose now that there is a distribution of N different initial conditions $\bar{\lambda}_i$, then

$$M(a,b) = (1/N) \Sigma_i A_I(a, \bar{\lambda}_i) \cdot A_{II}(b, \bar{\lambda}_i). \tag{4}$$

Guided by the previous result we write quite generally that every initial-value set during measurement of M(a,b) is given by $\bar{\lambda}_i = \bar{\lambda}_{L,i} + \bar{\lambda}_{NL,i}(a,b)$. Here $\bar{\lambda}_{L,i}$ is a purely local contribution, and $\bar{\lambda}_{NL,i}$ a small non-local correction that depends on a and b as a consequence of a long-range interaction between analysers and source. In the classic analysis by Bell the $\bar{\lambda}_{NL}$ are *identically* equal to zero. As just illustrated, every term in the above sum can be made at will equal to + or -1, depending on the values of the $\bar{\lambda}_{NL}$ (which do not need to be the same for I and II). Therefore by an ad-hoc choice of the N (>> 1) different $\bar{\lambda}_{NL}$ the sum can be made equal to cos2(a-b). For our model system we have explicitly verified by numerical calculation that this can be done through very small non-local corrections. One easily reproduces M(a,b) = cos2(a-b) by using $|\bar{\lambda}_{NL}|$-values that are $< 1.10^{-3}$ [12]. If one takes $t_m$ sufficiently long there is no difficulty (except for calculation time) to find cases in which these fluctuations are orders of magnitude smaller. This is due to the exponential growth of initial fluctuations in nonlinear systems with a positive Lyapunov exponent.

In fact it is possible to find deterministic systems for which the proof of the above thesis becomes quite straightforward. As an example, consider a nonlinear system with a separatrix dividing the phase space in two regions (say left and right), each region containing an attractor. Suppose that the system depends on some parameter x and that the separatrix is the same for all x in the interval $\Omega$. Take A = +(−) 1 if the trajectory is projected on the left (right) attractor. M is defined as in (4), with (a,b) $\in \Omega$. If the "local" initial values $\bar{\lambda}_{L,i}$ lie on the separatrix, then even vanishingly small departures from it will project the trajectory on the left or right attractor. Thus finite $\bar{\lambda}_{NL,i}$ make every term in the sum (4) equal to +1 or -1, depending on their position with respect to the separatrix. Therefore an adequate distribution of even arbitrarily small non-local $\bar{\lambda}_{NL,i}$ can make (4) equal to cos2(a-b), for all (a,b) $\in \Omega$. Notice that in this system a tight control over the measuring time of A is not required.

In the literature one has reported two "time-varying analyser" experiments [5,6] that were aimed at precluding non-local but Lorentz-invariant effects like we just considered. In these experiments relativistic separation would be obtained by measuring the correlation functions $M_i$ simultaneously with rapidly switching analysers. But in the first of these experiments [5] the changing of the analyser settings is periodic, and therefore it cannot exclude a broad class of Lorentz-invariant HVTs, as was pointed out in, e. g., Refs. [13,6]. Indeed, the analyser pair is periodically in the positions (a,b), (a',b), (a,b') and (a',b'), and thus some long-range interaction may induce a periodic fluctuation (with 4 "pseudo-periods") in some hidden variable $\lambda$. This could in turn imply that on average the $\lambda$ determining $M_i$ are different from those determining $M_j$, which is precisely the situation considered above ($\rho$ =

ρ(λ,a,b)). Therefore our previous analysis is applicable: if the polarization would sensitively depend on this hidden variable, then the quantum result of experiment [5] can be obtained by invoking a periodic signal with vanishingly small amplitude.

In a very recent experiment [6] the analyser settings are changed randomly and rapidly, and again the quantum predictions prove correct. Although also here some loopholes remain [14], this experiment comes remarkably close to the ideal EPRB scheme, and would seem to many physicists convincing in ruling out Lorentz-invariant HVTs. We will briefly show here, as a corollary, that it is trivial to formally extend the previous analysis also to this experiment. Consider a photon pair (contributing to the measurement of M(a,b)), and suppose that at the moment of impingement of a photon on its polarizer, some λ characterising the photon is a function of a *and* b. One can now introduce a distribution *at the analysers* of the form ρ = ρ(λ,a,b) to characterise the fluctuations in λ, and obviously the Bell-inequality can again be violated. If one allows for the possibility of a nonlinear evolution of λ in the finite [8] time interval Δτ between the impingement and the actual detection, then even extremely small non-local fluctuations can be dynamically enhanced and lead to the quantum result, as we have just illustrated [15]. Clearly, according to ref. [6] the non-local fluctuations characterised by ρ(λ,a,b) should now be considered as caused by a superluminal interaction. Even if the intensity of this interaction may, in principle, be arbitrarily small (compared to 1 newton), such an interaction would involve major changes in the present theoretical framework.

In conclusion, we have shown that the quantum result of the existing Einstein-Podolsky-Rosen-type experiments can be reproduced within deterministic models that include *arbitrarily* weak non-local effects. For the vast majority of existing experiments ([3-5]), these effects can be caused by (sub)luminal interactions. Since these interactions can, in principle, have an amplitude that is arbitrarily close to zero, they may have escaped from experimental investigation. Our result strongly corroborates the importance of genuinely loophole-free experiments of the type recently performed [6], as for such tests Lorentz-invariant interactions could not be invoked.

I am indebted to Prof. R. Omnès and Dr. V. Croquette for a critical reading of the manuscript. It is a pleasure and privilege to thank my colleagues Drs. M. Croci, W. De Baere, R. Ferrando, C. Fivez, A. Karlson, S. Métens, R. Morin, J. Tyvaert and P. Voisin for many stimulating discussions on the subject.

## APPENDIX

In this appendix -which is *not* included in the published article ! - I would like to discuss a few points in some more detail. It goes without saying that one can have several attitudes towards Bell's theorem. First of all: for those who are looking for some extra motivation for studying the subject, I would suggest to read a refreshing article written by N. D. Mermin [Physics Today, April 1985, p. 38]. Second, in the following (especially under point (iv)) I will have, occasionally, the hubris to stray out of the field of physics into philosophy. Some physicists may prefer to stop right here. Or in any case after point (iii). All philosophers will find the discussion under (iv) extremely limited.

Among the different attitudes one may adopt (in short: one has to choose between determinism and Lorentz-invariance), I would like to mention the following:

(i)     One considers that determinism is not an adequate principle. This is what the standard Copenhagen interpretation of quantum mechanics teaches us to do.

(ii)    One may consider that determinism is a stronger principle than Lorentz-invariance. After all, classical physics (and our daily world) seem to be governed by determinism: we have the impression that we can assign a cause to things that happen; nothing seems to happen without any cause. Or equivalently: if nature chooses one out of two equally possible events, then there should be a basis for this choice. One may thus believe that the results of the EPRB measurements (spin up or spin down) are predetermined and will be explained by future theories in a more complete manner. According to Bell and the results of the experiments[3-6] one has then to consider that Lorentz-invariance does not generally hold (!). Even if it has been one of the great principles of contemporary physics, it might prove inadequate for some yet to discover (microscopic) systems.

If one favours this attitude one may find in the present article an incitement to look for (very weak) superluminal interactions. Indeed, they can reconcile the measured quantum results with determinism. "Infinitesimally" weak superluminal interactions have of course a bigger chance to have escaped from experiments. It is an understatement to say that the investigation of such superluminal interactions is not a major field of research in contemporary physics.

(iii) One may hold fast on determinism and believe in the "loopholes" [3,7,8]. In other words one believes that the quantum results of the EPRB experiments will in the future be explained and completed with a Lorentz-invariant theory that uses one of these loopholes. The result of the present article seems to strengthen the pertinence of these loopholes: even if the effect considered in the loophole is initially extremely weak, it can be dynamically enhanced. In this context it may be interesting to study (by numerical simulation) experiments on complex nonlinear systems (for instance hydrodynamic systems, gases) in which one allows for the possibility of such a loophole.

(iv) A philosopher, and why not a physicist, may have the feeling that the ultimate solution of Bell's theorem is of logical order. It seems that this is the conclusion of many great thinkers, among which R. Feynman. N. D. Mermin ends his above mentioned article by quoting Feynman. He said about Bell's theorem: "You know how it always is, every new idea, it takes a generation or two until it becomes obvious that there's no real problem. I cannot define the real problem, therefore I suspect there's no real problem, but I'm not sure there's no real problem."

Above it is shown that even arbitrarily weak non-local effects can reconcile quantum mechanics and HVTs. From a logical point of view, one could argue that arbitrarily weak interactions can always be postulated. Then some people may find in the present article an argument to conclude that the alleged discrepancy between quantum mechanics and local HVTs is ... arbitrarily weak. This could be a nice illustration of Feynman's intuition that there is "no real problem" - or in any case that the problem is vanishingly small.

It goes without saying that a lot of adequate thinking needs to be done -for instance by a professional philosopher- before the ideas that I briefly and amateurishly expressed under (iv) may find some solid basis !

beating its wings in the Amazon forest, can cause a storm above Europe.